\documentclass[a4paper,aps,amsmath,amssymb,twocolumn,longbibliography,accepted=2023-05-19]{quantumarticle}
\pdfoutput=1
\usepackage[utf8]{inputenc}
\usepackage[english]{babel}
\usepackage[T1]{fontenc}
\usepackage{amsmath}
\usepackage{hyperref}

\usepackage{tikz}
\usepackage{lipsum}

\usepackage[utf8]{inputenc}

\usepackage{tikz}

\usetikzlibrary{decorations.markings}

\usepackage{amsmath}  % need for subequations
\usepackage{url}
\usepackage{hyperref}
\usepackage{mathtools}
\usepackage{siunitx}
\usepackage{multirow}
\usepackage{microtype}
\usepackage{array}
\usepackage{tabulary}
\usepackage{longtable}
\usepackage[version=4]{mhchem}
\usepackage{graphicx}
\newcolumntype{K}[1]{>{\centering\arraybackslash}p{#1}}

\usepackage{graphicx}   % need for figures
\usepackage{verbatim}   % useful for program listings
\usepackage{color}      % use if color is used in text
\usepackage{subfigure}  % use for side-by-side figures
\usepackage{hyperref}   % use for hypertext links, including those to external documents and URLs
\newcommand{\btxt}[1]{{\color{black} #1}}

\newcommand{\gtxt}[1]{{\color{black} #1}}
\newcommand{\dtxt}[1]{{\color{black} #1}}
\newcommand{\ctxt}[1]{{\color{black} #1}}

\newcommand{\nmtxt}[1]{{\color{black} #1}}
\usepackage{lipsum}

\begin{document}
%\title{Learnability of the dynamics of hard many-body problems:\\
%a case study on gap evolution in adiabatic quantum computing

\title{Deep recurrent networks predicting the gap evolution in adiabatic quantum computing}

\author {Naeimeh Mohseni}
\affiliation {Max-Planck-Institut f{\"u}r die Physik des Lichts, Staudtstrasse 2, 91058 Erlangen, Germany}

\affiliation{Physics Department, University of Erlangen-Nuremberg, Staudtstr. 5, 91058 Erlangen, Germany}

\author{Carlos Navarrete-Benlloch $^*$}
%\altaffiliation{Corresponding author: derekkorg@gmail.com tim.byrnes@nyu.edu}
\affiliation{Wilczek Quantum Center, School of Physics and Astronomy, Shanghai Jiao Tong University, Shanghai 200240, China}
\affiliation{Shanghai Research Center for Quantum Sciences, Shanghai 201315, China}
\affiliation {Max-Planck-Institut f{\"u}r die Physik des Lichts, Staudtstrasse 2, 91058 Erlangen, Germany}

\author{Tim Byrnes $^*$}
%\altaffiliation{Corresponding author: tim.byrnes@nyu.edu}
\affiliation{New York University Shanghai, 1555 Century Ave, Pudong, Shanghai 200122, China}
\affiliation{State Key Laboratory of Precision Spectroscopy, School of Physical and Material Sciences,East China Normal University, Shanghai 200062, China}

\affiliation{NYU-ECNU Institute of Physics at NYU Shanghai, 3663 Zhongshan Road North, Shanghai 200062, China}
\affiliation{Center for Quantum and Topological Systems (CQTS), NYUAD Research Institute, New York University Abu Dhabi, UAE}
%\affiliation{National Institute of Informatics, 2-1-2 Hitotsubashi, Chiyoda-ku, Tokyo 101-8430, Japan}
\affiliation{Department of Physics, New York University, New York, NY 10003, USA}

\author{Florian Marquardt}
\affiliation {Max-Planck-Institut f{\"u}r die Physik des Lichts, Staudtstrasse 2, 91058 Erlangen, Germany}

\affiliation{Physics Department, University of Erlangen-Nuremberg, Staudtstr. 5, 91058 Erlangen, Germany}

\begin{abstract}
\nmtxt{In adiabatic quantum computing  finding the dependence of the gap of the Hamiltonian as a function of the parameter varied during the adiabatic sweep is crucial in order to optimize the speed of the computation. Inspired by this challenge,} in this work we explore the potential of deep learning for discovering a mapping from the parameters that fully identify a problem Hamiltonian to the \ctxt{aforementioned parametric dependence of the gap} applying different network architectures. Through this example, we \gtxt{conjecture} that a limiting factor for the learnability of \dtxt{such problems} is the size of the input, that is, how the number of parameters needed to identify the Hamiltonian scales with the system size. We show that a long short-term memory network succeeds in predicting the gap when the parameter space scales linearly with system size. Remarkably, we show that once this architecture is combined with a convolutional neural network to deal with the spatial structure of the model, the gap evolution can even be predicted  for system sizes larger than the ones seen by the neural network during training. This provides a significant speedup in comparison with the existing exact and approximate algorithms in calculating the gap. 
\end{abstract}
\maketitle
%*****************************************************************************************

\section{Introduction\label{sec:introduction}}
\begin{figure}[h]
\centering
\includegraphics[width=1\linewidth]{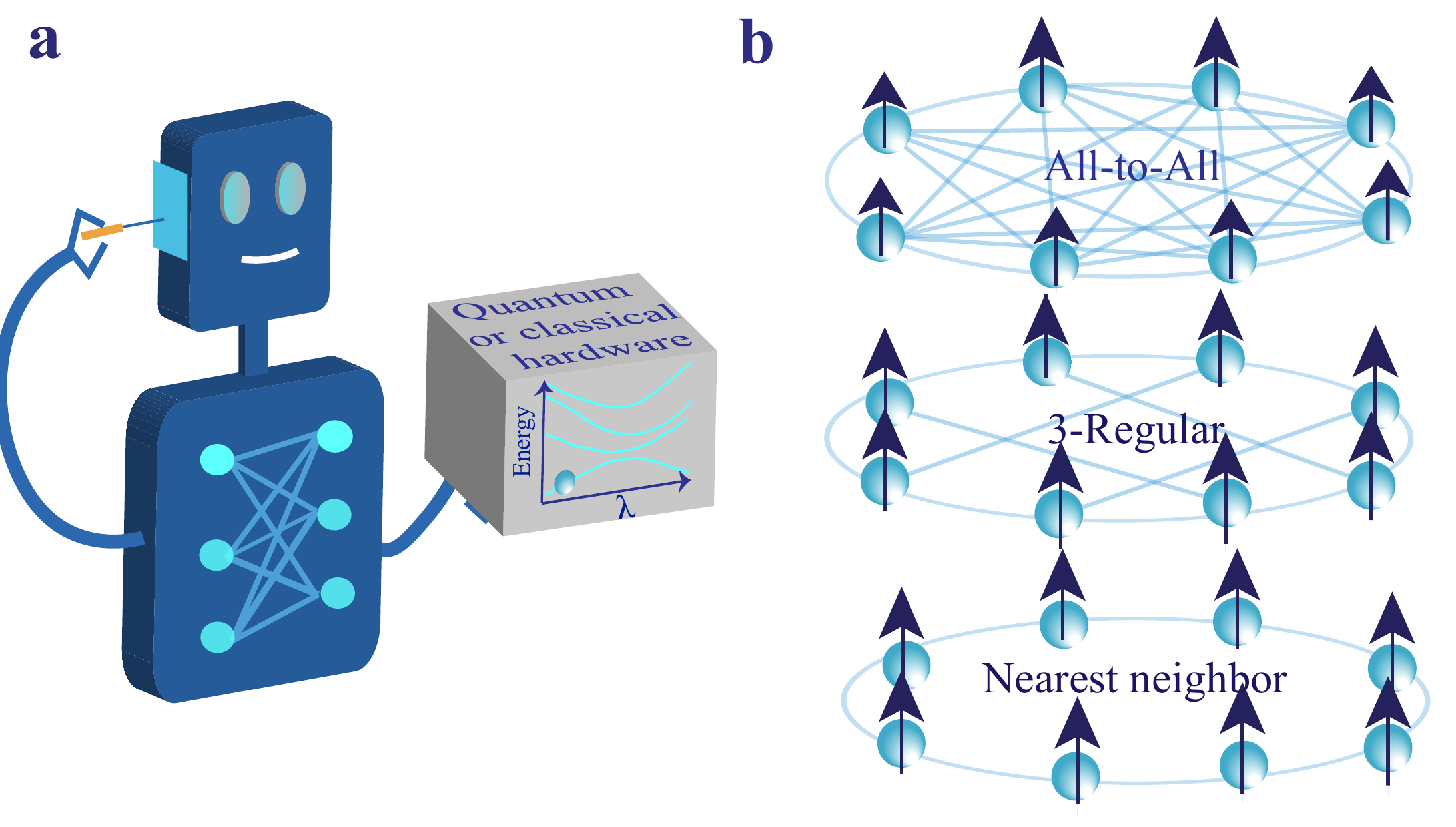} 
\caption {(a) A neural network learns the \ctxt{parametric gap} by observing the \ctxt{data} generated by a classical computer or a quantum computer. For the purposes of this work, we train the network on the data generated by a classical computer. (b) Spin models studied in this work that are taken to be the problem Hamiltonian in AQC. Spins are connected with random couplings.
\label{fig11}}
\end{figure}
Machine learning based on neural networks has demonstrated significant predictive capability for many challenging problems \cite{goodfellow2016deep,lecun2015deep}. In particular, its application in studying quantum many-body systems has attracted significant attention in the last few years \cite{carrasquilla2021neural}. A few notable successes include the use of neural networks in identifying quantum phases of matter and learning phase transitions \cite{van2017learning, PhysRevB.94.195105, wetzel2017unsupervised, carrasquilla2017machine, beach2018machine}, quantum state tomography \cite{ torlai2018neural, PhysRevX.8.011006}, enhancing quantum Monte Carlo methods \cite{PhysRevB.95.035105, bojesen2018policy}, solving optimization problems \cite{alam2020accelerating,medvidovic2021classical}, quantum control \cite{PhysRevX.8.031086}, the efficient representation of quantum states \cite{carleo2017solving, schmitt2019quantum}, and tackling quantum many-body dynamics \cite{schmitt2019quantum, mohseni2021deep,huang2021provably, vicentini2021machine}. 
%\btxt{In this work, we introduce a proficient  approach  in the latter direction.}

 Adiabatic quantum computing (AQC) is an example of a quantum many-body dynamical process and was proposed as an approach for solving optimization problems \cite{farhi2000quantum, mohseni2022ising}. \ctxt{Starting from the ground state of a reference Hamiltonian, a parameter of the system is varied adiabatically until reaching a target or problem Hamiltonian, whose ground state encodes the solution of the optimization problem.} % and has been proved as an alternative approach to the traditional gate-based quantum computing \cite{aharonov2008adiabatic}. 
 The sweep time in AQC for which adiabaticity can be achieved is proportional to a negative power of the minimum energy gap between the two lowest energy levels \ctxt{of the intermediate Hamiltonians that the system goes through during the sweep} \cite{roland2002quantum,lidar2009adiabatic,aharonov2003adiabatic}. Therefore, one way of optimizing the computation time is by spending the majority of the evolution time in the vicinity of \ctxt{any anti-crossings}. Creating such an annealing schedule requires prior knowledge of the \ctxt{dependence of the gap on the parameter that is swept, which we call the \textit{parametric gap},} and which is known to be as hard as solving the original problem \cite{amin2009first}. There are a handful of cases for which analytical expressions for the gap exist \cite{roland2002quantum, PhysRevA.78.032328, schutzhold2008dynamical, PhysRevA.73.022329}, but mostly the gap can only be determined numerically either based on exact diagonalization or approximate algorithms \cite{PhysRevLett.101.170503, mohseni2021error, altshuler2010anderson} which are limited to relatively small system sizes.  %Exact calculations are limited to the  system sizes smaller than 30 \cite{PhysRevA.71.062305}  and approximate algorithms are limited to the system sizes smaller than 128 \cite{PhysRevLett.101.170503}.
Therefore, in general there is no intuition about how the \ctxt{parametric} gap is related to the structure of the problem Hamiltonian.

%In this work we aim to study the learnability of the gap and explore the power of deep learning in discovering a mapping from the parameters that fully identify the adiabatic Hamiltonian to the full evolution of the gap. As a proof of principle we train the network on the data generated from numerical simulations but the final hope is that to train the network  than experimental data.  

\btxt{It remains an interesting open problem to characterize the complexity of learning \ctxt{the parametric gap} by applying deep learning, where one can think of two scenarios: training the network on either the data generated by a quantum computer or numerical simulations (Fig. \ref{fig11} (a)). In the former case, we have a hybrid quantum-classical algorithm. Such algorithms have recently attracted a lot of interest and have been examined in different contexts  \cite{niu2020learnability, mohseni2021deep,PhysRevLett.126.190505}, as  they can be applied for regimes \ctxt{for which} the run-time of the exact numerical simulations \ctxt{is} prohibitive\ctxt{, while} the network can \ctxt{still be} trained on the data generated via experiment. It is also interesting to  explore how the difficulty of the task is related to the complexity of the problem Hamiltonian and what are the models for which a neural network can assist in \ctxt{approximating the} gap.}

%One should note that in AQC  a highly precise estimation of the gap is not necessary as long as it is not overestimated significantly, since diabatic excitations are not produced as long as the adiabatic sweep time is longer than that required by the gap.

%Inspired by these open questions, here we explore the power of a neural network in discovering a mapping from the parameters that fully specify the adiabatic Hamiltonian to the evolution of the gap

\btxt{In this work, we address these open questions by exploring the performance of different network architectures trained on the \ctxt{parametric} gap for many different realizations of a random problem Hamiltonian. }% The gap dynamics can be generated either by  quantum or  classical hardware.
For the purposes of this work, we train the neural network on the data generated from numerical simulations rather than quantum hardware. However, our methods could equally be applied in a quantum-classical hybrid scenario, \btxt{given the availability of the desired hardware}.  
%
%Intuition says a hybrid quantum-classical scheme where a classical machine (network) is trained on data generated by a quantum computer  may  go beyond the classical machines for this task. As an approximate expression for the gap is already helpful in order to reduce the AQC computation time. Recently, such hybrid algorithms applying neural networks have attracted interests and have  been examined  in different contexts . 

%
As for the problem Hamiltonian in AQC, we consider spin models as test cases.     We  find that LSTM networks, which are typically used for sequence processing and prediction, including audio signal analysis \cite{graves2013speech} and language translation \cite{mikolov2010recurrent} excel in predicting the gap evolution in certain scenarios.
These architectures are known to be good at capturing both long-term and short-term dependencies in sequences. This characteristic is extremely useful as it gives the LSTM network the power to handle complex dynamics. Remarkably, we demonstrate combining LSTM with a convolutional network  called  CONVLSTM \cite{xingjian2015convolutional}  allows extrapolation to larger system sizes in some settings. This architecture combines a convolutional neural network (CNN) to deal with the spatial structure of the input  with the LSTM  that tracks the time evolution. 
 
 While throughout this work we concentrate on the particular task of gap prediction in AQC, our study provides insight into more general many-body dynamics. We conclude that an important limiting factor for the learnability of such dynamics is the way in which the number of parameters is needed to identify the Hamiltonian scales with the system size. % Moreover, our study demonstrates the promise of  convolutional recurrent neural networks for extrapolating the dynamics of inhomogeneous time-dependent quantum many-body models beyond the system sizes that they are trained on.

\section{Problem definition  \label{NN}}
In this section, we define the  models that we explore in this work. We also introduce the neural network architectures that we apply and explain how we train them. 
\subsection{Model}

We consider the AQC Hamiltonian defined as
\begin{equation}
H\ctxt{(\lambda)}=\ctxt{(1-\lambda)} H_{0}+\ctxt{\lambda} H_{\textrm{p}},
\end{equation}
with
\begin{subequations}
\begin{align}
H_{\textrm{p}}&=\sum_{i,j=1}^{M}  J_{ij} \sigma_{i}^{z} \sigma_{j}^{z}+\sum_{i=1}^{M} K_{i} \sigma_{i}^{z},\label{Ising} \\
H_{0}&=-\sum_{i=1}^{M} \sigma_{i}^{x},
\end{align}
\end{subequations}
where $\sigma_{i}^{\alpha}$ with $\alpha=x,z$ are Pauli operators, and $J_{ij}$ and $K_i$ are random coefficients that identify the problem Hamiltonian. \ctxt{In practice, AQC considers a time-varying parameter $\lambda$ that is swept from 0 to 1, following some schedule $\lambda(t)$. However, this specific function of time is irrelevant for our purposes, since} our goal is finding a mapping from the parameters $(J_{ij},K_i)$, which we collect in a matrix $\boldsymbol{J}$ and a vector $\boldsymbol{K}$, to the \ctxt{gap between the ground and first excited states of $H(\lambda)$ parametrized as a function of $\lambda$}. \dtxt{In the following we denote this parametric gap by $g(\lambda)$.} We study in particular two limiting cases \ctxt{for the} problem Hamiltonian:  i) all-to-all connected spin models and ii) 1D nearest neighbor connected models and \gtxt{iii) 3-Regular graphs shown in Fig.  \ref{fig11} (b)}. Note that all parameters are dimensionless since they are normalized to the energy scale of $H_0$, which we take as the reference. 

\dtxt{Let us emphasize that while we are here considering the problem of parametric gap prediction in AQC, the same approach can be applied to find a mapping from the (possibly time-dependent) parameters of any many-body Hamiltonian, to the temporal dynamics of some observable. One simply has to train the network on trajectories of the observable of interest, taking time as a parameter that would play the role of $\lambda$ in our AQC example.}

%The system is initially prepared in the ground state of $H_0$ and the Hamiltonian gradually transitions to the desired problem Hamiltonian $H_{\textrm{p}}$. Based on the adiabatic theorem if one performs the sweep sufficiently slowly, the system will remain in its instantaneous ground state throughout the evolution \cite{farhi2000quantum}. 

%We believe our scheme  can succeed in predicting other interesting concepts for local models that can be identified with parameters that scale at most linearly with system size. For example, the $1D$ Ising model despite being a simple model plays a key role in an intuitive understanding of many principles and phenomena in the context of many-body and statistical physics. Therefore,  our scheme supports, along with previous efforts, that  neural networks as an instrumental tool can be applied for an intuitive comprehension  and a rapid exploration of  features of  quantum many-body models. 
\begin{figure}{h}
\centering
\includegraphics[width=0.9\linewidth]{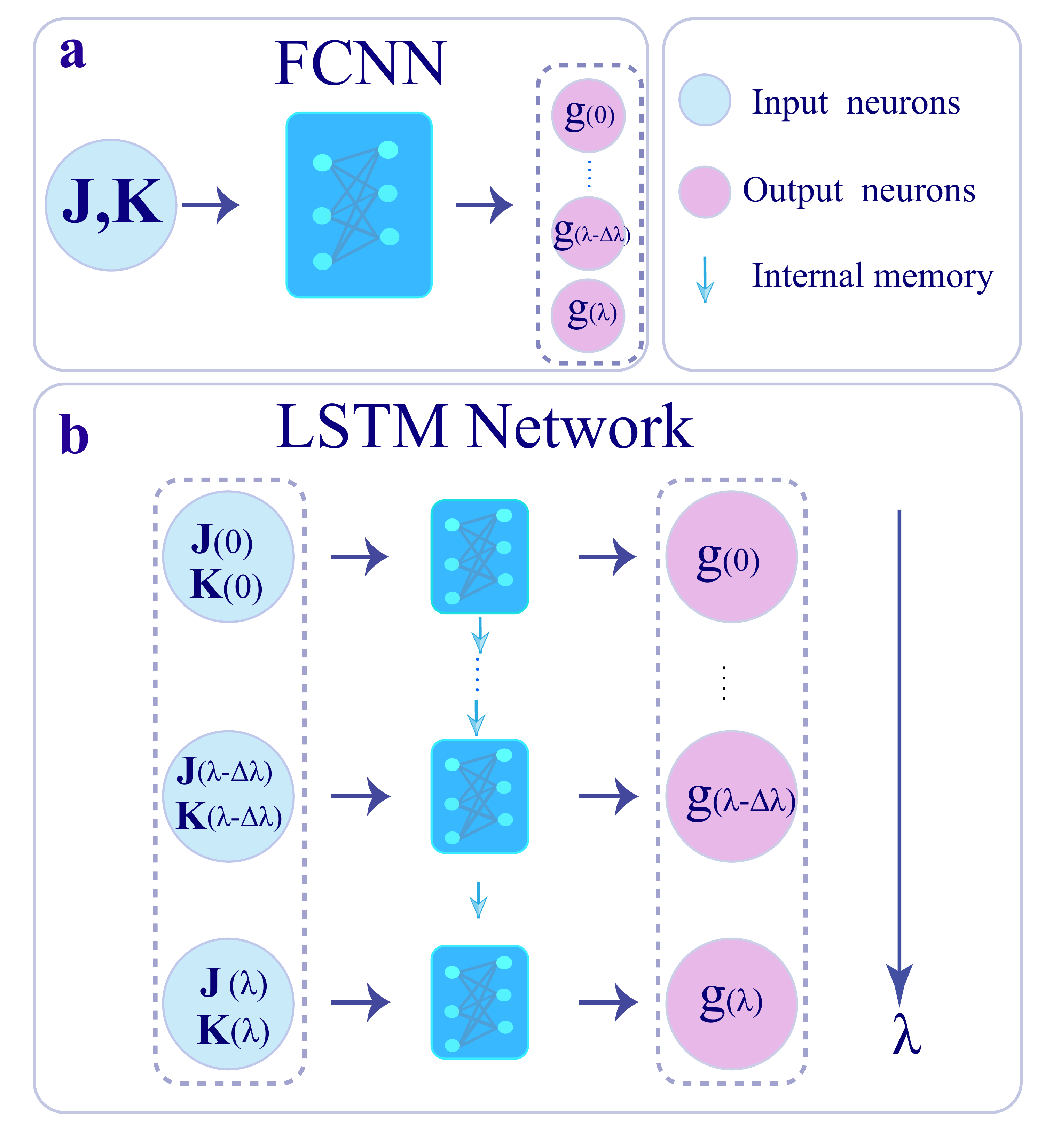} 
\caption {Schematic representation of the neural networks considered in this work. (a) Fully connected neural network (FCNN)  with  the bare parameters $\boldsymbol{J}$ and $\boldsymbol{K}$ as the input and the \ctxt{parametric gap as} the output. (b) The long short-memory (LSTM) network with $\boldsymbol{J}(\ctxt{\lambda})=  \ctxt{\lambda}\boldsymbol{J}$ and $\boldsymbol{K}(\ctxt{\lambda})= \ctxt{\lambda}\boldsymbol{K}$ as the input and the gap as the output denoted by $g(\ctxt{\lambda})$. Arrows  between the modules (dark blue cells) indicate the content of the internal neural memory being passed \ctxt{from a given value of $\lambda$ to the next considered value, which we take $\Delta\lambda$ apart}.
\label{fig1}}
\end{figure}
\subsection{Neural network architectures}  
We apply three neural network architectures for our goal:   an FCNN, an LSTM network, and a CONVLSTM network.  For the FCNN architecture,  we feed into the neural network the parameters $ \boldsymbol{J} $ and $\boldsymbol{K}$ that fully identify the problem Hamiltonian and the neural network  provides as output the \ctxt{parametric gap }(Fig.~ \ref{fig1} (a)), which we signify by
\begin{equation}
   (\boldsymbol{J},\boldsymbol{K}) \xrightarrow{\text{FCNN}} g(\ctxt{\lambda}).
\end{equation}

In the LSTM architecture, the input is not just the bare parameters that identify the problem Hamiltonian, but the effective contribution of these parameters during the whole sweep, namely $\ctxt{\lambda}\boldsymbol{J}$ and $\ctxt{\lambda}\boldsymbol{K}$ (Fig. \ref{fig1} (b)). These coefficients effectively identify the full adiabatic Hamiltonian during the sweep. In this case, then we have
\begin{equation}
   (\ctxt{\lambda} \boldsymbol{J},\ctxt{\lambda}\boldsymbol{K}) \xrightarrow{\text{LSTM}} g(\ctxt{\lambda}).
\end{equation}

The most important difference between these two architectures is that the LSTM receives as input the effective contribution of the parameters \ctxt{for each value of $\lambda$ up to the one it wants to compute}, eventually working its way sequentially through the whole \ctxt{$\lambda\in[0,1]$ interval}. In contrast, the FCNN has to work out the full \ctxt{parametric gap} at once. In addition, as shown in Fig.~\ref{fig1}(b), the LSTM architecture is composed of modules (dark blue cells). Each module is made of a few gates (see Sec. II of the Supplemental Material for more details) which  decide on the flow of information in and out of each module at each time.

To study the possibility of extrapolating the predictions to system sizes beyond what the neural network has been trained on we also apply  a CONVLSTM  network \cite{xingjian2015convolutional}. This architecture is designed  for data with spatio-temporal input \cite{shi2015convolutional}.  It combines a CNN to deal with the spatial structure of the input with the LSTM that tracks the \ctxt{``evolution'' in $\lambda$}. CNNs  can be applied for variable input size therefore helping to scale up the predictions to larger sizes \cite{goodfellow2016deep,nielsen2015neural} (See Supplemental Material Secs. I and II for more details). Inspired by this feature and using an appropriate preparation of the input such that we can present properly the spatio-temporal structure of the input to the network, we explore the power of the CONVLSTM in predicting the gap for larger system sizes than that it has been trained on. The input and the output of the network in this case are signified as  % The internal matrix multiplications are also exchanged with the convolution operations.
\begin{equation}
   (\ctxt{\lambda}\boldsymbol{J}(x),\ctxt{\lambda}\boldsymbol{K}(x)) \xrightarrow{\text{CONVLSTM}} g(\ctxt{\lambda}),
\end{equation}
where $\boldsymbol{J}(x)$  and $\boldsymbol{K}(x)$ should properly present the spatial structure of the problem Hamiltonian as we explain in more detail in Sec. \ref{extrapolation}.
 
\subsection{Training} 
To train the neural network, we generate a set of random parameters $\boldsymbol{J}$ and $\boldsymbol{K}$ for the particular system size of interest. We then diagonalize the Hamiltonian \ctxt{for the desired values of $\lambda$} and calculate the first two eigenvalues, \ctxt{whose difference provides the parametric gap}. %at  30 points (evenly distributed) for the particular sweep time $\tau=30$. 
All these parameters are taken from a uniform distribution in the interval $[-1,1]$.  To improve the performance of the neural network at regions where the gap is smaller,  we train it on the $\text{log}(g(\ctxt{\lambda})+1)$ rather than $g(\ctxt{\lambda})$. %We show that then the network is able to predict the gap for any unseen distribution the parameters, e.g., a binary choice of $\lbrace-1,1\rbrace$ or a Gaussian distribution.

Out of the generated random instances, we keep a set to  evaluate the neural network, which we call the test set, and a set for validation. The validation set is used to  fine-tune the hyperparameters of the neural network, but no training occurs on this set. The remaining instances are used for training. %Later, we discuss how the performance of our networks relates to the training set size. 
To evaluate the performance of the neural network, we calculate the mean square error $\textrm{MSE}$ on our test set defined as
\nmtxt{\begin{equation}
\textrm{MSE}=\langle |\text{log}(g_{\text {true}}(\ctxt{\lambda})+1) -\text{log}(g_{\text {predict}}(\ctxt{\lambda})+1)|^{2}\rangle _{n,\ctxt{\lambda}},
\end{equation}}
where the $g_{\text {true}}$ and the $g_{\text {predict}}$ denote the true gap calculated by diagonalizing the Hamiltonian and the predicted gap obtained by the neural network, respectively. The average is taken over the  number of instances $n$ and the parameter $\ctxt{\lambda}$. % We train the network on the logarithm of the gap rather the gap itself to improve the performance of the network at the regions that the gap is smaller. 

\begin{figure*}[t!] 
\centering
\includegraphics[width=1\linewidth]{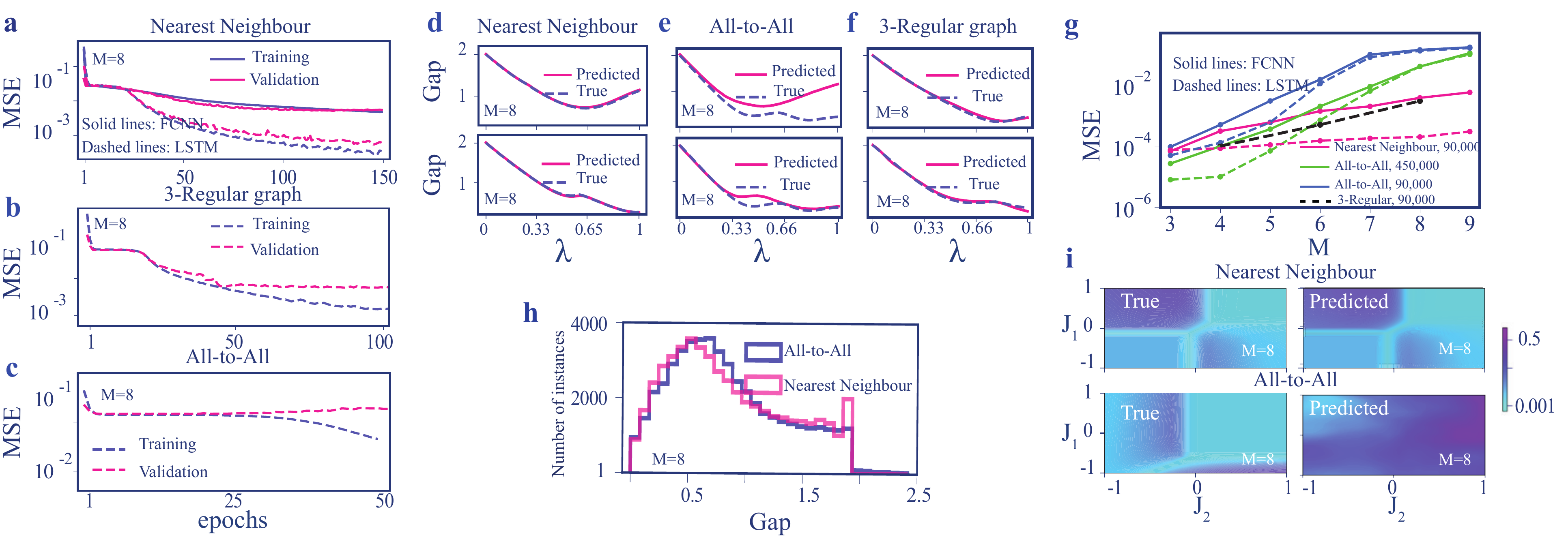} 
\caption {Comparing the performance of the LSTM network and FCNN  in  predicting the \ctxt{parametric gap} for the all-to-all model, the nearest neighbor connected model, \gtxt{ and 3-regular graph}.  Validation $\textrm{MSE}$  versus epochs for system size $M=8$ applying the FCNN (solid lines)  and  LSTM network (dashed lines) for (a) the nearest neighbor, \gtxt{ (b) 3-regular graph}, (c) all-to-all connected models. 90,000 samples are used to train the neural network and 10,000 samples are used for the validation.  Predicted gap (solid line) and exact gap (dashed line) in terms of $  \lambda$ with $\tau=30$ for a few typical problem instances for the (d) nearest neighbor connected model, (e) all-to-all connected model,  \gtxt{and (f) 3-regular  graph}.   (g) The error $\textrm{MSE}$ versus system size on the test set (containing 1000 samples for each system size) for all models applying both FCNN (solid curves) and LSTM network (dashed curves). The numbers on the legend denote the training set size. The shown points are average over 5 attempts for training.  (h) Histogram of the gap during  the whole  sweep for both the nearest neighbor and all-to-all connected models. (i) The true and predicted minimum gap versus couplings $J_1$ and $J_2$, where all other couplings and local terms are fixed. 
 \label{fig2}}
\end{figure*}

%b) Predicted gap versus real gap for the second half of the evolution for all-to-all connected model with different system sizes. Color denotes the $\textrm{MSE}$.

\section{Gap evolution learnability}

In this section, we study the learnability of the gap by applying the different neural network architectures that we discussed above. %We also explore the possibility of predicting the gap to the system sizes beyond what the network is trained on.
We  first study the performance of our neural network architectures in predicting the gap for spin models in two extreme limits: either all spins are connected or only nearest neighbors in 1D 
shown in (Fig. \ref{fig11} (b)). \gtxt{In addition, we  study the learnability of the gap for 3-regular graphs.} We inspect how the prediction accuracy scales with system size for  these models applying our  different network architectures. 
%In order to explore the power of our networks in predicting the gap evolution, we identify the main bottlenecks for the learnability of the gap and inspect how the prediction accuracy scales with system size. 

%We consider adiabatic duration $\tau=30$ and we calculate the gap in 30 evenly distributed points.
In Fig. \ref{fig2}(a), (c), we show the $\textrm{MSE}$ over training and validation sets in terms of epochs for the nearest neighbor and all-to-all connected models.  The number of epochs indicates the number of times that the network has seen all the training instances. For the nearest neighbor connected model, it can be seen that for both LSTM network and FCNN  the error decreases with the number of epochs. However, it is evident that the LSTM network performs better (Fig. \ref{fig2}(a)).   In contrast, for the all-to-all connected model, the network overfits (Fig. \ref{fig2}(c)). For this model, we just showed the results using LSTM network as we observed the same behavior applying the FCNN as well.
Investigating different factors such as the network size and the training set size, we learned that the latter is the main bottleneck in this case. 

In Figs.~\ref{fig2}(d) and (e), we show the predicted and the true gap for a few typical instances applying the LSTM network. It is clear that our network fails to predict the gap for the all-to-all connected instances while successfully  predicts the gap with high accuracy for  the nearest neighbor-connected instances.

%To study more systematically the failure of the network  for the all-to-all connected model,  in Fig. \ref{fig2} (c) we show the predicted minimum  gap versus the true minimum gap for different system sizes. 
%The plot for each system size contains 200 problem instances. The color denotes the $\textrm{MSE}$ error in predicting the gap throughout the full evolution. It is evident that for larger system sizes than $M=7$, the predicted values start to deviate from the true values for most of the instances.  This is due to the fact that  100,000 samples seems to be insufficient for the network to learn the gap above $M=7$. One can see a mixed performance for the network on the system size $M=7$. It seems  there are still cases, sitting on the diagonal, for which the network succeeds in predicting the gap. For these cases on the diagonal the color is more bluish meaning that for these rare cases that network succeeds in predicting the minimum gap it  can also predict the full gap trajectory successfully. 

In Fig.~\ref{fig2} (g), we study how the error scales with the system size for a fixed number of training samples specified in the legend. The shown errors are found by averaging over 1000 test instances. Let us focus first on the nearest neighbor-connected model. It is obvious that the LSTM network has a considerably higher precision and the error in prediction scales more favorably in terms of system size for this architecture in comparison with the FCNN (compare the pink solid line with the pink dashed line). We attribute this to the fact that the LSTM architecture has memory and is able to record both long and short-term dependencies. This architecture has the built-in notion of causality, while the FCNN needs to learn causality on its own as it has no notion of time. Applying the same number of training samples for the all-to-all connected model (blue lines), the error grows more dramatically with system size in comparison with the  nearest neighbor-connected model. The error can be decreased by increasing the training set size, but even a factor of 5 (green lines) is not enough to achieve reasonable accuracy for larger system sizes (say $M>6$). This implies that the number of samples required to train the network explodes with the system size when aiming at a given error in the predictions. As a consequence of this explosion, the LSTM network does not seem to show a better performance  in comparison with the FCNN for $M>5$ in the figure when 90,000 instances are used for training (compare dashed and solid blue lines). 

We have explored some reasons why the network fails for the all-to-all connected model, while it succeeds for the nearest neighbor model when for a given system size $M$ the Hilbert space dimension is the same for both models. We conjecture the main reason is  due to the way in which the parameter space size scales with the system size for each model: linearly for the nearest neighbor model and quadratically for the all-to-all connected one. Note that by parameter space size we mean the number of parameters that identify the model. One may think that another reason might be that the nearest neighbor connected model can be also simpler in terms of the gap size and complexity of the gap trajectories. To investigate this, we compare the gap size during the sweep for both the nearest neighbor and the all-to-all connected models for system size $M=8$ in Fig.~\ref{fig2} (h). \ctxt{The plot is made for one random instance of each model, but similar figures are found for other instances. It is apparent that, in general, the typical size of the gap is similar for both models. Ideally, one would wish to investigate whether the gap in the fully connected case is more feature-rich compared with the nearest neighbor case by analyzing the dependence of the minimum gap with all parameters $\boldsymbol{J}$ and $\boldsymbol{K}$. Unfortunately, this is computationally unfeasible. Therefore, we consider a simpler analysis, in which we analyze the dependence of the minimum gap with two of the parameters only, e.g. $J_1$ and $J_2$, for different random instances of all other parameters. For all random instances, we find a dependence similar to that shown in Fig.~\ref{fig2} (i).  As can be seen, the pattern of the minimum gap in the all-to-all connected model does not appear to be more complicated than the nearest neighbor connected model. However, the network fails in predicting the minimum gap for the all-to-all connected model. This seems to further support the idea that the all-to-all connected model is not more feature-rich than the nearest-neighbors one.} Another potential reason is that the local nature of the connectivities in the  nearest neighbor model can make it easier for the network to learn the dynamics of the gap. If that were to be true, then one should expect an improvement by encoding the all-to-all connected model into a local model. In Sec.~\ref{LHZ}, we introduce such an encoding through the so-called Lechner-Hauke-Zoller (LHZ) mapping \cite{lechner2015quantum}, and find the same scaling of the error with the system size.

\gtxt{To further explore the validity of our conjecture --- that the success of the network is connected with the way that parameter space size scales with system size --- we also studied 3-regular graphs for which each spin is connected to three other spins. In Fig. \ref{fig2}(b), we show the MSE over training
and validation sets in terms of epochs for the LSTM network. In Fig. \ref{fig2}(f), we show the predicted and true gap for two typical instances.  In Fig. \ref{fig2}(g), the black curve  shows the performance of our LSTM network in predicting the gap for this model where for each system size we averaged over 1000 realizations. In overall, the precision in learning the gap  on this model is lower in comparison with the nearest neighbor connected model. Nevertheless,  the network is able to make predictions with reasonable accuracy.  The reason for the lower precision is that for the 3-regular graph, the number of connectivities scales as $ 3M/2 $  while it scales as $M$ for the nearest neighbor connected model. Therefore, to get comparable accuracy for the predictions, one needs to provide more samples to train the 3-regular graph model.} 

\nmtxt{Note that our conjecture that the learnability of the network is connected with the parameter space size, i.e. the number of connectivities, is consistent with well-known aspects of optimization problems.  For optimization problems, the complexity of the problem increases by increasing the density of the problem,  namely increasing the connectivities of the problem Hamiltonian \cite{pittel1996sudden, biroli2002phase, ozaeta2022expectation}.}

As a last remark of this section, one may argue that it might be easier for the network to learn the times for which the gap becomes small, rather than the \ctxt{full parametric dependence of the gap}. We have also investigated this and observed that the network still fails for the all-to-all connected model. Our conjecture is that this is again a consequence of the quadratic scaling of the parameter space size with the system size.

\section{Extrapolation \label{extrapolation}}
\begin{figure*}[t!]
\centering
\includegraphics[width=1\linewidth]{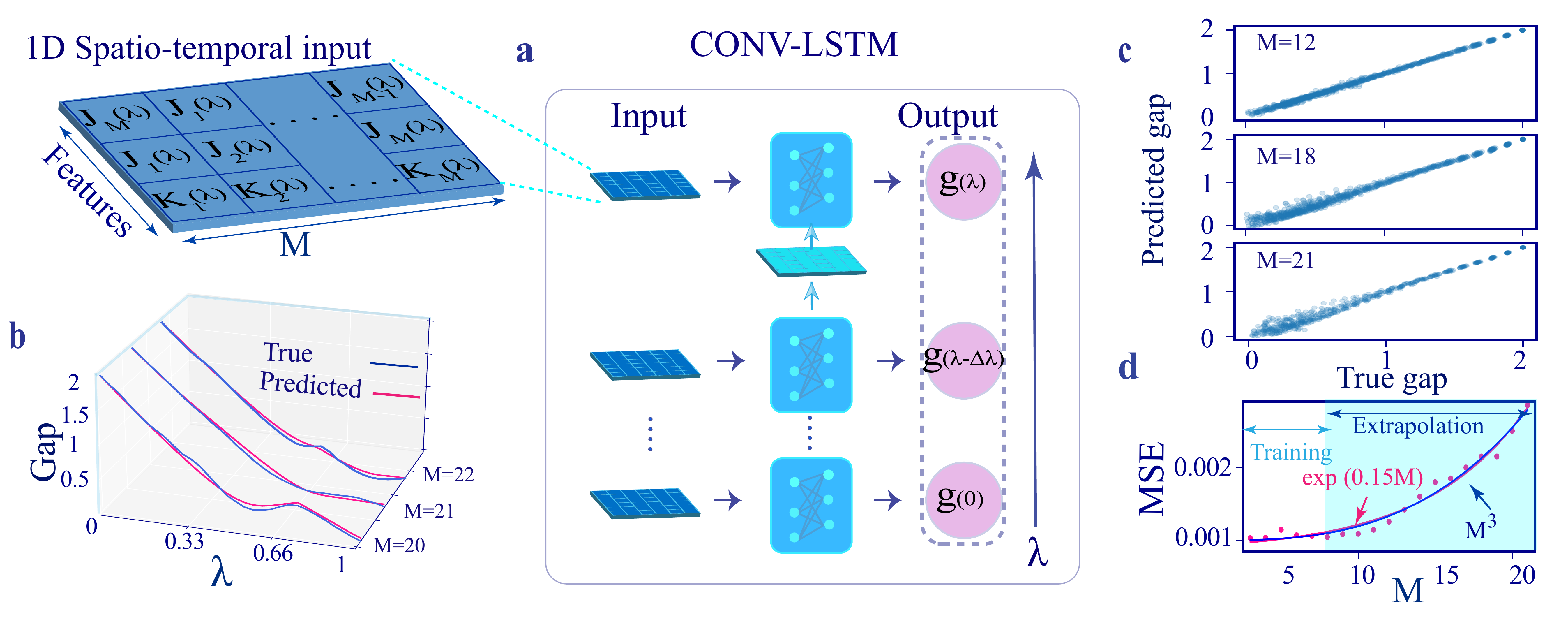} 
\caption{Extrapolating the gap to the larger system sizes applying 1D-CONVLSTM network for the nearest neighbor connected model. (a) Schematic representation of the CONVLSTM with  spatio-temporal inputs (light blue sheets). Arrows between the modules (dark blue cells) indicate the content of the internal neural memory being passed \ctxt{one value of $\lambda$ to the next $\lambda+\Delta\lambda$} and the dark blue sheet denotes a convolutional structure for module-to-module transitions. %  The blue sheet at the top left shows structure of the 1D spatial input. 
(b) Predicted  (solid line) and true gap (dashed line) in terms of $\lambda$ with $\Delta\lambda=1/30$ for three typical problem instances.  (c) Predicted gap versus the true gap \ctxt{for all values of $\lambda$ and 20 random problem instances for each system size}. (d) The $\textrm{MSE}$ error in predicting the gap versus system size. Network is trained on system sizes with $M \in[3,9]$ and is evaluated on test samples with $M\in [3,22]$. The shown errors are averaged over 200 random problem instances for each system size.  The highlighted region marks the system sizes that the network has not been trained on.  90,000 samples are used to train the network.
 \label{fig3}}
\end{figure*}

In this section, we explore the possibility of the network to predict the gap for system sizes beyond those in which it is trained. As indicated already, the potential architectures for this goal are CNNs, which can be applied for an input with variable size.  These architectures are  known to be good for extracting features on local models. Therefore, we apply to the nearest neighbor connected model the CONVLSTM architecture which  is designed for sequence prediction problems with spatial structure.  This network extends the fully connected LSTM architecture to have a convolutional structure in both the input-to-module (dark blue sheet Fig.~ \ref{fig3}(a)) and module-to-module (light blue sheet Fig.~ \ref{fig3}(a)) transitions.

%The filter is a matrix that moves over the input data, performs the dot product with the sub-region of input data, and gets the output as the matrix of dot products

 %

In Fig.~ \ref{fig3}(a), we show the input of the network with a 1D spatial structure and containing three features at each site. The first feature represents the corresponding local parameter $\boldsymbol{K}$ at each site. The second and third ones represent the couplings $\boldsymbol{J}$, arranged so as to emphasize that each site is connected to its two nearest neighbors. See Supplemental Material Sec. III for more details on the technical implementation and the layout of the network.

 We train the network on system sizes $M\in[3,9]$, and then evaluate it on the test samples with system sizes $M\in[3,22]$. We observed that our CONVLSTM network succeeds in predicting the gap both for the system sizes that it has been trained on and larger system sizes than that. In Fig. ~\ref{fig3} (b), we show the predicted gap (solid line) and the true gap (dashed line) as a function of $\ctxt{\lambda}$ for three typical problem instances with $M=20,21,22$. As can be seen, the network is able to approximate the gap during the evolution with good precision. We also find that there is a correlation between the size of the gap and the accuracy of the network. In Fig. ~\ref{fig3} (c), we show the true gap against the predicted gap \ctxt{for 20 test instances and all considered values of $\lambda$}. It is clear that close to the regions where the gap is smaller, the performance of the network is less accurate, even on an absolute level.

 In Fig.~ \ref{fig3} (d),  we show how the prediction error in prediction scales with the system size. The highlighted region marks the system sizes that the network has not been trained on.  We have not been able to conclude whether the error scales polynomially or exponentially, since both functions fitted relatively well.

 %Rather than looking at an entire image at once to find certain features it can be more effective to look at smaller portions of the image. Each layer in a CNN is made up of a number of channels, each channel can be regarded as a feature detector working to detect a specific feature.The presence of each of these features would be encoded in the activation map of each channel. So they share parameters (weights and biases), as each channel only maintains a single set of parameters, these greatly reduces the number of parameters. This means that all the neurons in a given convolutional layer respond to the same feature within their specific response field. which have no limitations on the input size at all because once the kernel and step sizes are described, the convolution at each layer can generate appropriate dimension outputs according to the corresponding inputs. %diverse and interesting uses %highly effective at classifying structured data where the order of arrangement matters including images, audio and video

In the previous section, we observed for the all-to-all connected model, using 90,000 samples for training, the network is able to learn the gap on small system sizes with $M<7$. We are interested to study whether a CONVLSTM network that is trained on that small sizes can still extrapolate the predictions to the larger system sizes.  
To apply the CONVLSTM network for the all-to-all connected model, we need to map it to a local model. This is because CNNs are good at extracting features on the local models as neurons of different layers are locally connected. In the next section, we map the all-to-all connected model to a local model and explore the potential of CONVLSTM for extrapolating the predictions to larger system sizes.

\section{Mapping the all-to-all connected model to a local model \label{LHZ}}
\begin{figure*}[ht]
\centering
\includegraphics[width=1\linewidth]{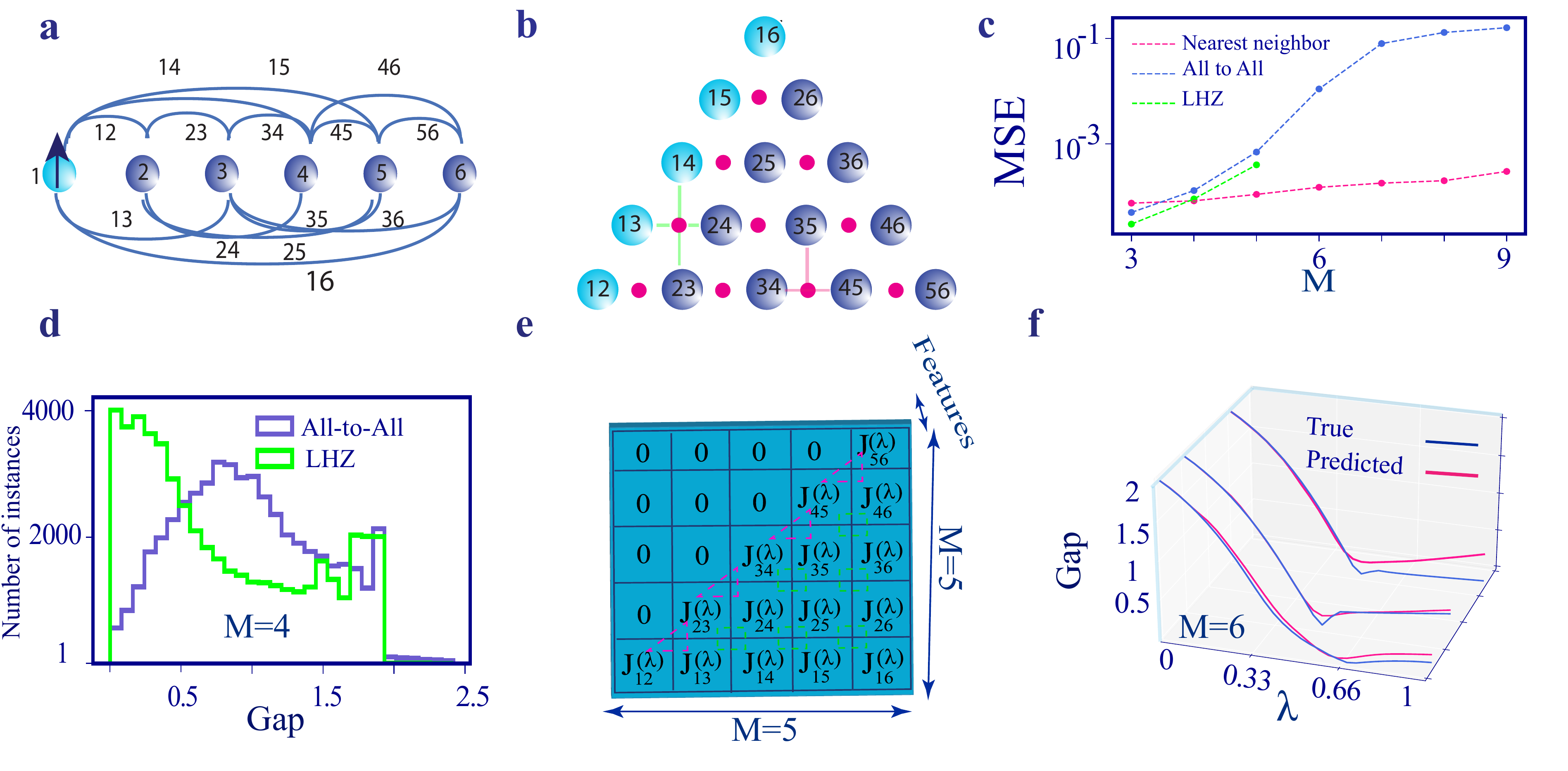} 
\caption {  The power of the LSTM network in  predicting the \ctxt{parametric gap} for the LHZ model with $C=3$. (a) An all-to-all connected model with system size $M=5$ where  the light blue circle  shows an auxiliary logical qubit (fixed to 1)  to implement single-body terms. (b) The LHZ architecture  where the light blue circles show physical qubits that encode single-body terms. The four qubits around each small pink circle (plaquette) consist  of four-body constraints. (c) The error for predicting the \ctxt{parametric gap} averaged over 1000 realizations for all three investigated models versus system size applying LSTM network.  90,000 instances are used  to train the network for all the shown models and system sizes. (d) Histogram of the gap during the sweep for  the all-to-all connected and LHZ models for system size $M=4$. (e) Spatio-temporal input of the 2D-CONVLSTM containing one feature at each site which presents the couplings. The couplings are arranged on the shown square such that they respect the connectivities in (b). The dashed squares and three-angles denote connections between qubits that identify the three-body and four-body constraints on LHZ model. The square at the top left shows the second feature  which identifies for the network where a qubit exists.   (f) Exact gap and predicted gap (applying  2D-CONVLSTM) in terms of $\lambda$ with $\Delta\lambda=1/30$ for three particular problem instances with size $M=6$ where the network has not been trained on. 
 \label{fig4}}
\end{figure*}  

  In this section, we first investigate whether the highly non-local nature of the all-to-all connected model is  one of the reasons that makes it hard for the network to predict the gap. To understand this, we apply the LHZ mapping  \cite{lechner2015quantum} to encode this model into a model with only local connectivities and check if the network performs better. LHZ maps the graph of the $M$ all-to-all connected logical qubits (Fig.~ \ref{fig4} (a))  onto a planar graph with $N_p=M(M-1)/2$ physical qubits and only local connectivities (Fig.~ \ref{fig4} (b)). Within this mapping, the original problem Hamiltonian (\ref{Ising}) can be encoded into the following Hamiltonian in the physical qubit basis
 \begin{equation}
 H_{\mathrm{p}}^{\mathrm{LHZ}}=\sum_{k=1}^{N_p} J_{k} \sigma^{z}_{ k}-C\sum_{\langle i, j, k, l\rangle}  \sigma^{z}_{i} \sigma^{z}_{ j} \sigma^{z}_{ k} \sigma^{z}_{ l},\label{eqLHZ}
 \end{equation}
  where $\langle i, j, k, l\rangle$ denotes sum over nearest neighbor spins. Energy penalties in the second term which involve $N_p-M+1$ four-body interactions (the four qubits around each small pink circle shown in Fig.~ \ref{fig4} (b)) are introduced to \ctxt{ensure that in the $C\rightarrow\infty$ limit the low-energy sector of $H_{\mathrm{p}}^{\mathrm{LHZ}}$ reproduces the spectrum of the original Hamiltonian (in practice, we have checked that $C=3$ suffices to observe the convergence of the ground state in all the cases we have studied).} In the bottom row only three qubits are connected, for which three-body constraints are introduced instead. 
  
  Single-body terms in the original model, which correspond to a local field acting on the logical qubits, can be also implemented by adding an  auxiliary logical qubit (light blue qubit in Fig.~ \ref{fig4} (a) ) fixed to state $|0\rangle$, which is an eigenstate of $\sigma_z$ with eigenvalue +1. Interaction of this  auxiliary logical qubit with the rest of the qubits implements these local terms, which correspond to $M$  extra physical qubits (light blue qubits in Fig.~ \ref{fig4} (b)) in the LHZ architecture \cite{lechner2015quantum}.

  %Now let us illustrate the performance of our NN for gap prediction applying LHZ model for the system sizes that we have been able to provide samples with exact calculation of gap using solving Schrödinger equation. We show the MSE in terms of number of epochs on both validation and training set applying deep FFNN built of dense layers and LSTM-DNN in Fig \ref{LHZ}(c) for system size $M=4$ (correspond to 10 physical qubits). To train both architectures, 90,000 samples are applied. Comparing MSE on validation set with training set, it is evident that  NN is able to learn the gap quite well. 
%We illustrate the performance of the network on system sizes that we  have been able to provide samples for the network.
In Fig.~ \ref{fig4} (c), we compare the performance of the LHZ model with the originally all-to-all connected model applying the standard LSTM network. 90,000 instances are applied to train the network for each model and system sizes up to $M=5$, for which the LHZ model already contains 15 physical qubits. We also show the results for the nearest neighbor connected model for the purpose of comparing the scaling. \btxt{ We observe that for the LHZ model the error scales with the system size in a similar way to the original model. While we have not been able to analyze for larger $M$ values, if locality played a strong role, we see no reason why it should not be visible even for small system sizes. Therefore overall, these results suggest that the locality of the connectivities should not improve the way that error scales with the system size.} %but overall error decreases with a prefactor. %As the improvement is not really significant, it remains unclear whether this improvement is a consequence of local nature of the connectivities.  Due to the numerical limitations,  we have not been  able to train the network for sufficiently large system sizes to understand this better. Our impression from current results is that the locality of the connectivities should not play a significant role.
 %The reason is that, in LHZ model the random coefficients appear behind local terms and coefficient of four-body constraints ($C$)  that involves nearest-neighbor interactions is fixed. This coefficient  should be normally chosen sufficiently large so that the lowest energy levels of the original model and LHZ model to be similar.
  
  In Fig.~ \ref{fig4} (d), we compare the typical size of the gap for both models. As can be seen, the LHZ model has substantially more instances with smaller gap sizes as compared to the all-to-all connected model, but still, the network achieves comparable or even better precision. This confirms our conjecture that the main bottleneck is not the size of the gap, but rather the way in which the parameter space size scales with the system size. %Of course, we expect at the end the regions for which gap is smaller are in general more challenging for the network. 

 Now we explore whether mapping the all-to-all connected model to a local model and applying a CONVLSTM network helps to extrapolate the predictions to larger system sizes.  For this case, we need a 2D-CONVLSTM as the LHZ model has a $\textrm{2D}$ spatial structure. In Fig.~ \ref{fig4} (e), we show the spatio-temporal input of our 2D-CONVLSTM for the case with $M=5$.  Each site contains one feature which represents the coefficients  $\boldsymbol{J}$. These couplings are arranged on the shown square such that they respect the connectivities in the LHZ model.   Dashed squares and triangles  mark  qubits around each small pink circle in Fig.~ \ref{fig4} (b). 
 %For the same reason that we explained in the previous section, we need to pad our input with additional cells to provide an opportunity for cells in border to interact with filter. Therefore,  to clarify for the network where there exist a qubit as for the second feature we assign one to the cells that qubits are placed and zero otherwise. 
 We train the network on system sizes $M\in\{3,4,5\}$ and evaluate it on system size $M=6$ (Fig.~ \ref{fig4} (e)), which is the largest size (includes 21 physical qubits) that we are able to generate a few samples for. As can be seen, the network is still able to  predict the gap for a larger system size that it has not been trained on, but not with a high accuracy such as in the nearest neighbor connected model. Due to the numerical limitations, we have not been able to evaluate our network for larger system sizes, but considering the low precision for $M=6$, we expect the network to fail for the larger sizes. We attribute this partially again to the fact that the quadratic scaling of the parameter space with system size necessitates exponentially growing resources for training. However, we expect our 2D CONVLSTM succeeds in predicting the dynamics of 2D models for which the parameter space size scales more favorably with the system size.
 
\section{Speedup} 

%Our scheme by predicting the full gap evolution for certain models may provide speedup in two ways.  First in the context of AQC, knowing the spectral gap helps to create a highly optimized schedule for annealing. Thereby, reducing the computation time in AQC.  Second, our network is able to predict the gap  with considerably higher speed in comparison with both existing exact and approximate algorithms which is useful for other applications that may require knowledge about the gap. 

In this section, we discuss the potential speedup that can be achieved by employing the network for the nearest-neighbor model, for which we have shown the network succeeds in making predictions and in extrapolating. 

Supervised training of neural networks can be motivated in many ways, e.g. an explicit algorithm to turn the input into the desired output may not even be known in principle (as in image classification). However, for a scenario like the one discussed here, a numerically exact algorithm exists. One can then think of at least three reasons why training and deploying a network might still be beneficial: (i) the run-time of the network is so much shorter than the run-time of the exact algorithm, and we will use it for prediction on so many problem instances, that it is worth to invest the numerical effort needed to generate a large number of training examples in the first place; (ii) the run-time of the algorithm scales very unfavorably with system size and the network is able to make reasonably accurate predictions also for larger sizes, even if it was not trained on those; (iii) there is a regime where the run-time of the exact algorithm is prohibitive, but data may be generated in another way (e.g. via experiments), and the network can still be made to learn in this regime and be used for prediction. We will discuss the last point (training from experiments) in the following subsection and focus on the numerical speedup here.  

A real advantage will be obtained if the eventual number of problem instances $N_{\rm use}$ that we intend to apply the network to is sufficiently large. Specifically, if $N_{\rm train}$ is the number of training samples we needed to achieve good accuracy, we have to fulfill the inequality 

$$ N_{\rm train} (\tau_{\rm Alg}+ \tau_{ \rm train}) + N_{\rm use} \tau_{\rm NN} < N_{\rm use} \tau_{\rm Alg} ,$$
where $\tau_{\rm Alg}$ is the run-time of the algorithm itself and $\tau_{\rm NN}$ is that of the neural network, for one problem instance. %This inequality presumes that the bottleneck for network training is the training data generation by the algorithm; otherwise the total amount of network training time $\tau_{\rm train}$ spent per one training sample during all epochs also needs to be taken into account and added to $\tau_{\rm Alg}$ on the left-hand-side.
$\tau_{\rm train}$ denotes training time spent per one training sample during all epochs. If a network trained on small systems can also be applied to larger system sizes, where the algorithm run-time becomes $\tau_{\rm Alg,Large}$, we need to insert that larger value on the right-hand-side, making application of the neural network more favorable (even if its own run-time might also increase somewhat).

% In general, the achieved speedup depends on the algorithm that the network replaces. However, the ability of our network to predict the gap for larger system sizes that it has never seen suggests that the gained speedup may go much beyond possible speedups achieved by any exact or approximate  numerical simulations. 
With this in mind, let us provide some illustrative numbers with all the caveats regarding the dependence on computer hardware and algorithm. We trained the network on small system sizes, which is inexpensive, taking just a few hours to generate $N_{\rm train}=90,000 $ samples, plus about one day to train the network. The largest system size for which we have been able to generate  a few tens of instances to evaluate the neural network is $M=22$. For this size, it takes one and a half hours to calculate the gap evolution for a single instance when applying the Lanczos algorithm. In contrast, once the neural network is trained, it is able to predict the gap evolution for this system size in 5ms. Assuming a scenario with these illustrative numbers, we can use the formula above to conclude that the application of the neural network would become favorable if it were deployed on $N_{\rm use}>20 $ actual problem instances which is notably less than the number of training samples, due to the performance gains via extrapolation. Any application on more instances would yield strong time (and memory) savings in comparison to direct use of the algorithm.

\section{ Hybrid algorithm \label{implemetation}}
%Our protocol can be adapted to training the neural network on larger system sizes from data generated by a quantum annealer. In this section we comment on the cost of such a hybrid implementation and argue about its near-term feasibility. We emphasize that there are a lot of caveats for such implementation, and we point out some of them in this section. Therefore, the following ideas should be taken as raw estimations. 

\nmtxt{Our protocol is adaptable for training neural networks on larger system sizes using data generated from a quantum annealer. In this section, we discuss the near-term feasibility and associated costs of a hybrid implementation. We caution that there are caveats to consider and highlight. In general, measuring the gap using a quantum annealer is currently not straightforward. However, there are proposals for achieving this in the future. Therefore, the ideas presented in this section should be considered rough estimations and only applicable when such proposals are implemented.}

\ctxt{Let us start commenting on the feasibility and cost of measuring the gap in experiments. This can be done by} applying different methods. Spectroscopic techniques are widely used for this purpose, for example using Ramsey-like interferometry \cite{matsuzaki2021direct, russo2020evaluating}. In all methods\ctxt{, for each value of $\lambda$ we want, one needs to measure an observable by repeating the experiment $n$ times, resulting in an error that scales as $\sim 1/\sqrt{n}$, set by the projection noise}. Including the number $N$ of samples that we need to prepare for training, overall we then need $n \ctxt{N_\lambda} N$ runs, where \ctxt{$N_\lambda$ stands for the number of values of $\lambda$ that we take}. The number of training samples depends on the system size $M$ we want to train the neural network on. Since in this work we have trained the network on sizes up to $M=9$, it is not easy to see a clear scaling for the number of samples required to achieve a given accuracy. Therefore, we roughly estimate $N\sim 2 \times 10^5$ from our experience assuming one trains the network on the problem instances with $M=50$. Then, taking $N_\ctxt{\lambda}\sim 50$ and $n\sim 10^4$, one would require around $10^{11}$ runs. This implies that it seems feasible to train the network on data generated from quantum annealers implemented on platforms for which each run takes up to a few microseconds, which would then require about one day to generate data to train the network. This is indeed the case for superconducting circuit platforms \cite{albash2018demonstration}, which \ctxt{provide a leading experimental candidate where our ideas should be feasible}.

In the future, it will be an interesting challenge to explore how much a network can deal with more noisy measurement data, reducing the need to accumulate statistics. Some efficiency improvements are possible, e.g. one might allow for a larger statistical error (smaller $n$) if the time points are closely spaced because the network will then effectively try to interpolate smoothly through the noisy observations.

 \section{Conclusion and outlook\label{conclusion}}
 In this work,  we explored the power of deep learning in discovering a mapping from the parameters that fully identify a problem Hamiltonian to the \ctxt{parametric gap of the corresponding AQC Hamiltonian}. We observe the CONVLSTM  network succeeds in predicting the gap on  models for which the parameter space size scales linearly with the system size. %This suggests our scheme is applicable also for certain type of regular graphs that fulfil this criteria.In a regular graph each vertex has the same number of connectivites. 
For such models,  our CONVLSTM network is even able to predict the gap for system sizes larger than that it has been trained on  and may provide speedup in comparison with the existing exact and approximate algorithms. While during this work we concentrated on the gap prediction in AQC,  our study can provide insight for more general many-body dynamics.  We \gtxt{conjecture} that one of the limiting factors for the learnability of such dynamics applying supervised learning is the way that parameters identifying the model scale with the system size.

Our study supports the promise of CONVLSTM networks in predicting the dynamics of inhomogeneous  many-body systems and their potential for extrapolating the dynamics beyond what the neural network is trained on \cite{PhysRevE.102.033301, mohseni2023deep, blania2022deep}.  Our scheme can also be applied in the context of quantum approximation optimization algorithms  as it can be viewed  as a trotterized version of AQC  with parametrized annealing pathway. \gtxt{In this context, it can in particular be integrated  with ``divide and conquer''  methods  where one splits problems with hundreds of qubits to sub-problems with a few tens of qubits and   the dynamics of desired observables of each sub-problem can be learned and predicted by the network \cite{guerreschi2021solving, saleem2021quantum, PRXQuantum.3.010346}. }

\nmtxt{We acknowledge the publication of a paper \cite{hegde2022deep} that recently confirmed our research findings in this work}.

 % Overall, our scheme supports the previous efforts \cite{mohseni2021deep,carrasquilla2021neural} demonstrating the capability of neural networks as an instrumental tool for  intuitive comprehension and  rapid exploration of quantum many-body dynamics. 

\begin{acknowledgments}
N.M  would like to thank the Erwin Schrödinger International Institute for Mathematics and Physics(ESI), University of Vienna (Austria) for the opportunity to participate in the Thematic Programme "Quantum Simulation - from Theory to Application" where part of this work has been accomplished and for
the support given. T. B.  is supported by the National Natural Science Foundation of China (62071301); State Council of the People’s Republic of China (D1210036A); NSFC Research Fund for International Young Scientists (11850410426); NYU-ECNU Institute of Physics at NYU Shanghai; the Science and Technology Commission of Shanghai Municipality (19XD1423000); the China Science and Technology Exchange Center (NGA-16-001). CNB acknowledges sponsorship from the Yangyang Development Fund, as well as support from a Shanghai talent program and from the Shanghai Municipal Science and Technology Major Project (Grant No. 2019SHZDZX01).
\end{acknowledgments}

$^*$ \textbf{Corresponding authors:}  derekkorg@gmail.com,  tim.byrnes@nyu.edu

%*****************************************************************************************
\setcounter{equation}{0}
\setcounter{figure}{0}
\setcounter{table}{0}
\setcounter{section}{0}
\setcounter{subsection}{0}

\makeatletter
\renewcommand{\theequation}{S\arabic{equation}}
\renewcommand{\thefigure}{S\arabic{figure}}

\begin{widetext}

   \begin{center}
\textbf{\large Supplemental Material}
	\end{center}

	In this Supplemental Material, we provide a brief review of the convolutional neural network and a particular type of convolutional recurrent neural network called  convolutional long short-term memory.  We also  provide  details related to the layout of the network architectures that we applied. 
\end{widetext}

\maketitle

\section{Convolutional neural networks}
\nmtxt{Convolutional neural networks (CNNs)  are specific types of networks with a grid-structured topology \cite{goodfellow2016deep}. They are composed of multiple layers of neurons, including convolutional layers, pooling layers, and  sometimes fully connected layers. The convolutional layers apply a series of filters (also known as kernels) to the input image, each of which is responsible to detects specific features. The pooling layers then downsample the output of the convolutional layers by taking the maximum or average of small groups of neurons, reducing the dimensionality of the data and improving computational efficiency.  A convolutional neural network (CNN) uses data that contains a few features at each point of the spatial grid. For instance, a 2D-CNN takes an input with shape $(w,h,c)$ where $w$ and $h$ represent the height and width of the spatial structure of the input, and $c$ represents the number of features at each point of the spatial grid. In our case, the features at each site of the grid are identified by the coefficients that describe the spatial structure of our problem Hamiltonians, as shown in Fig. 3 (a) and Fig. 4 (e).} . %The kernel size should be chosen such that it respects the type of the locality that exist in the model or  the way that each site is affected by its neighbor sites. For example, for the nearest neighbor connected model we choose the kernel size 3 as each site is connected to its two nearest neighbors. 

CNNs have built-in affine invariance so they can recognize patterns that are shifted or tilted in the input. One known benefit of CNNs is that they can be applied for input with varying spacial structures helping  to scale up  the predictions to larger sizes. Due to this feature, they do not use the standard matrix multiplication but convolution instead. These are the main features that we exploited in this work to extrapolate the prediction of our network  to the large sizes. The type of affine invariance and the locality that is required for these architectures to make best of them all exist in our nearest neighbour or the encoded version of the all-to-all connected models. 
 \section{Convolutional Recurrent Neural Networks}

In this section, we provide a brief review of the recurrent neural networks (RNNs) and a particular type of  that called long short-term memory (LSTM).  Then we explain  an extended version of that called convolutional LSTM (CONVLSTM) network which we applied in this study.  

RNNs are built of a chain of repeating modules of neural networks. Such a network introduces a feedback loop such that the output of the network at the current time depends on the current input ($x_t$), called the external input, and also on the perceived information from the past, called the hidden input ($h_{t-1}$) \cite{nielsen2015neural}. \btxt{Note that in our AQC example $\lambda$ plays the role of time in this architecture. } Such a network is able to record the history for -- in principle -- arbitrary long times, since weights are not time-dependent and therefore the number of trainable parameters does not grow with the time interval. For training such a network, the gradient of the cost function needs to be backpropagated from the output towards the input layer, as in feedforward networks, and also along backward along the time axis.

However, RNNs are prone to run into a fundamental problem, the ``vanishing/exploding gradient problem'', i.\,e., that the size of the gradient decreases (or sometimes increases) exponentially during backpropagation. In principle, this problem can also occur in traditional feedforward networks, especially if they are deep. However, this effect is typically much stronger for RNNs since the time series can get very long. This seriously limited the trainability of early RNN architectures, which were not capable of capturing long-term dependencies. This problem led to the development of RNNs with cleverly designed gated units (controlling memory access) to avoid the exponential growth or vanishing of the gradient, and therefore permitting to train RNNs that capture both long and short-term dependencies. The first such architecture is called LSTM, developed by Hochreiter and Schmidhuber in the late 90s \cite{hochreiter1735long}. % and allows the RNN to keep learning over many time steps
%This structure has been refined by many people in following work.
\begin{figure}{}
	\centering
	\includegraphics[width=1.0\linewidth]{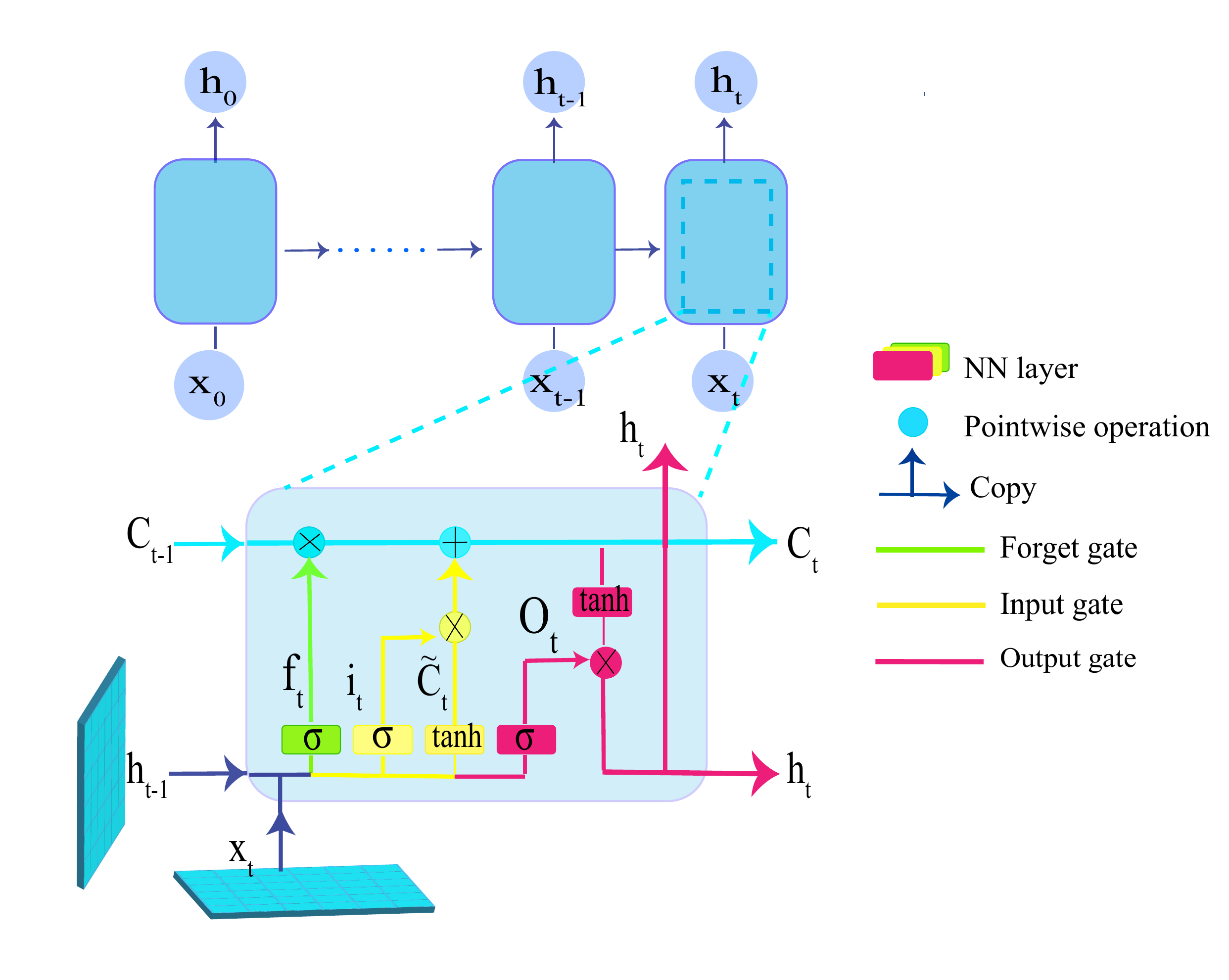}
	\caption{An CONVLSTM network made of a chain of repeating modules, where each module includes three gates. Both the input-to-state and module-to-module transitions have a convolutional structure.}
	\label{figs1}
\end{figure}
As an RNN architecture, standard LSTM is also built of a chain of repeating modules, as is shown in Fig. \ref{figs1}, where the repeating modules have a more complicated structure than in a simple recurrent  network. Each module includes three gates, where each gate is composed out of a sigmoid network layer, together with the point-wise multiplication on top of it. Next, we explain step by step how these three gates together control how the memory needs to be accessed. We label weights $w$ and biases $b$ by subscripts according to the name of the corresponding layer.

\begin{itemize}
	\item{Forget gate layer: this gate uses the hidden state $h_{t-1}$ from the previous time step and the external input $x_t$ at a particular time step $t$ (with the bias $b_f$ and the weight $w_f$) to decide whether to keep the memory, or to discard the information that is of less importance, applying a sigmoid activation.
		\begin{equation}
			f_{t}=\sigma\left(W_{f} \cdot\left[h_{t-1}, x_{t}\right]+b_{f}\right).
		\end{equation}
		$\sigma$ denotes sigmoid function and the dot stands for matrix multiplication. Eventually, the output of the forget gate is multiplied with the module state ($C_t$).}
	\item{Input gate layer: the operation of this gate is a three-step process,
		\begin{itemize}
			\item{first a sigmoid layer decides which data should be stored (very similar to the forget gate) %and then that will be eventually added to the state
				\begin{equation}
					i_{t} =\sigma\left(W_{i} \cdot\left[h_{t-1}, x_{t}\right]+b_{i}\right).
				\end{equation}}
			\item{hidden state and current input also will be passed into the tanh function to push values between -1 and 1 to regulate the network and stored in $\tilde {C}_t$.
				\begin{equation}
					\tilde{C}_{t}=\tanh \left(W_{C} \cdot\left[h_{t-1}, x_{t}\right]+b_{C}\right).
				\end{equation}}
			\item{ the outcome of the two previous steps will be combined via multiplication operation and then this information is added to the module state ($f_{t} * C_{t-1}$).
				\begin{equation}
					C_{t}=f_{t} * C_{t-1}+i_{t} * \tilde{C}_{t}.
				\end{equation}}
		\end{itemize}
		Here the $*$ denotes element-wise multiplication.}
	\item{Output gate layer: the operation of this gate which decides the value of the next hidden input can be decomposed into two steps,
		\begin{itemize}
			\item{run a sigmoid layer on the previous hidden state and the current input, which decides what parts of the module state are going to be carried
				\begin{equation}
					\begin{array}{l}
						O_{t}=\sigma\left(W_{o}\left[h_{t-1}, x_{t}\right]+b_{o}\right). \\
					\end{array}
				\end{equation}}
			\item{passing the module state through tanh to squish the values to be between -1 and 1, and finally multiply it by the output of the sigmoid gate so that we only pass to the next module some selected parts
				\begin{equation}
					h_{t}=O_{t} * \tanh \left(C_{t}\right).
				\end{equation}}
		\end{itemize}}
\end{itemize}
 
When data besides temporal structure have also spatial structure the extended version of this architecture called CONVLSTM can be applied. In this case  
 there is convolutional structure in both the input-to-module and module-to-module transitions \cite{shi2015convolutional} (shown with blue sheets in Fig. \ref{figs1} ) and the internal matrix multiplications are exchanged with the convolution operations.

%One advantage to convolutional networksis that they can also process inputs with varying spatial extents. These kinds ofinput simply cannot be represented by traditional, matrix multiplication-basedneural networks. This provides a compelling reason to use convolutional networkseven when computational cost and overﬁtting are not signiﬁcant issues.

%Convolutional networks provide a way to specialize neural networks to workwith data that has a clear grid-structured topology and to scale such models tovery large size. This approach has been the most successful on a two-dimensionalimage topology.

% Each convolutional layer is composed of multiple filters. Having less weights allows Convolutional NN to converge faster to a minima, while reducing the probability of overfitting and using less memory and taking less training time.

\section{Neural networks layout}
In this section we present the  layout of  the architectures that we applied for the gap prediction task. We have specified and trained all these different architectures with Keras \cite{chollet2015keras}, a deep-learning framework written for Python. \nmtxt{We systematically tested different values for hyperparameters for  different architectures that we applied here.  For example for activation function, learning rate, batch size, the kernel size of the convolutional network, etc. After conducting several rounds of experiments, we identified the combination of hyperparameters that produced the best results on our validation set.  We report these parameters below.} 
\subsection{Fully connected  neural network}
 In Table. \ref{table_1}, we summarize the details related to the layout of our fully connected  neural network (FCNN) for different system sizes and all the models that we explored. The training set size for all models and system sizes is 90,000.
Since  for the all-to-all connected model, the number of samples required for training  explodes with the system size, the network fails in predicting the gap for $M>7$ using 90,000 samples for training.

We use the rectified linear activation function ($\mathrm{ReLU}$) for all layers except the final one, which utilizes a linear activation function. Additionally, we use the popular ``adam" optimizer with a default learning rate of 0.001.

\begin{table} [!h]
\scalebox{0.8}{
\begin{tabular}{| *{7}{c|} }
    \hline
FCNN    & \multicolumn{2}{c|}{All-to-All}
            & \multicolumn{2}{c|}{NN}
                    & \multicolumn{2}{c|}{LHZ}
                                     \\
    \hline
System size   &   \# HL &   \#N/L  &   \# HL  &   \#N/L  &   \# HL  &   \#N/L   \\
    \hline
M=5   &   5  &   500  &   5  &   500  &   3  &   500   \\
    \hline
M=6  &    7   &   700    &  5     &    500   &    4   &  500            \\
\hline M=7 &\multicolumn{2}{|c|}{Fails}& 6 &700   &- &-\\
\hline M=8 & \multicolumn{2}{|c|}{Fails}&6 &700   &- &-\\
\hline M=9 & \multicolumn{2}{|c|}{Fails}& 6 &700 &- &- \\
    \hline
\end{tabular}}
\caption{The layout of the FCNN  for different system sizes. $\#$  HL and $\#$ N/L denote the number of hidden layers and the number of neurons per layer, respectively.  }
\label{table_1}
 \end{table}

\subsection{Convolutional long short-term memory}
In this section, we present the layout of the 1D and 2D CONVLSTM architectures that we applied in the main text. % Let us point out here that the main important difference between an LSTM architecture with a FCNN is that the LSTM receives as input only the information for a single time step and outputs the predicted observables for that time step, eventually working its way sequentially through the whole time interval.

\textbf{1D-CONVLSTM}
  In Table. \ref{table_2}, we present the layout of our 1D-CONVLSTM network applied in Sec. IV.  CONVLSTM layers  capture the temporal-spatial dependencies of the input.  TimeDistributed  is a wrapper that applies a layer to every temporal slice of an input.  We use this wrapper together with the global max pooling  to transfer the input with  the temporal-spatial structure to the output with temporal structure.

%Dense layers kill the translational invariance, therefore making the network fail to extrapolate the predictions to larger sizes. To overcome this problem,
To push the network to succeed in extrapolation, we did a pre-processing of the input data, as we explain next. Assume that we want to train the network on sizes $M \in [3,M_T]$ with $N$ training samples for each system size, finally evaluating it on $M \in[3,M_E]$, with $M_E>M_T$ the largest extrapolation we explore. We prepare a four-dimensional array with size $N(M_T-3)\times N_t\times M_E\times 3$ filled with zeros. Here $N(M_T-3)$ is the total number of training samples, $N_t$ denotes the number of time steps, and 3 denotes the number of features as we explained in the main text. Now for each sample and time step, we place the string with $M_T$ coupling coefficients randomly within the $M_E$ zeros in this third dimension of the array.% We found out that this helps the network to succeed in extrapolating.

\begin{table} [!h]
\begin{tabular}{|l|l|l|l|}
\hline Layers& Filters & Kernel size\\
\hline CONVLSTM1D&20 & 3  \\
\hline CONVLSTM1D& 40& 3  \\
\hline CONVLSTM1D& 60 & 3 \\
\hline CONVLSTM1D& 40 & 3  \\
\hline CONVLSTM1D& 20 & 3  \\
\hline  \multicolumn{3}{|c|}{TimeDistributed(Global max pooling)} \\
\hline
\end{tabular}
\caption{The  layout of the 2D-CONVLSTM}
\label{table_2}
\end{table}  

Note that in a CNN by default, a filter starts at the left of the input with the left-hand side of the filter placed on the far left pixels of the input. The filter is then stepped across the input until the right-hand side of the filter is placed on the far right pixels of the image. This means the edge of the input is only exposed to the edge of the filter. However,  starting the filter outside the frame of the image can give the pixels on the border of the image more of an opportunity for interacting with the filter and therefore more of an opportunity for features to be detected by the filter, and in turn, an output feature map that has the same shape as the input image. This needs to add the border of input some pixels which is called padding. But to make it clear for the network where exactly the edge of the input starts we add a row of features that are filled with ones for the range of pixels that input starts and ends identifying where a qubit exists and where not.

\textbf{2D-CONVLSTM}
In Table.~ \ref{table_3}, we present the layout of our 2D-CONVLSTM network applied in Sec. V. We prepare the input of this network also with the same pre-processing instruction that we explained for our 1D-CONVLSTM.  The only difference is that here we need to insert randomly a square array with size ($M_t-1$, $M_t-1$) inside the input array of the network with size ($M_e-1$, $M_e-1$).

\begin{table} [!h]
\begin{tabular}{|l|l|l|l|}
\hline Layers& Filters & Kernel size\\
\hline CONVLSTM2D&20 & (2,2)  \\
\hline CONVLSTM2D& 40& (2,2) \\
\hline CONVLSTM2D& 60 & (2,2)\\
\hline CONVLSTM2D& 40 & (2,2)  \\
\hline CONVLSTM2D& 20 & (2,2) \\
\hline  \multicolumn{3}{|c|}{TimeDistributed(Global max pooling)} \\
\hline
\end{tabular}
\caption{The  layout of the 2D-CONVLSTM applied in Sec. III C.}
\label{table_3}
\end{table} 
\nmtxt{\section{Training data }
As we pointed out in the main text we train the network on small system sizes $M<8$. To generate data for training, we use qutip \cite{johansson2012qutip}. Regarding the time steps for which to calculate the gap, we empirically inspected the gap behavior for different system sizes under different resolutions. From our observations, we found that a ratio of $\frac{g(\lambda+ \Delta \lambda)-g(\lambda)}{g_{Max}}<0.05$  is  a good choice. We acknowledge that this number may not be suitable for all scenarios.} 
\bibliography{paper}

\end{document}